\documentclass[11pt]{article}

\usepackage{amsfonts}
\usepackage{amsbsy}
\usepackage{amssymb}

\usepackage[pdftex]{graphicx}

\newcommand{\BE}{\begin{equation}}
\newcommand{\EE}{\end{equation}}
\newcommand{\half}{{\textstyle{\frac{1}{2}}}}
\newcommand{\msbar}{\overline{\rm MS}}
\newcommand{\lamR}{\tilde{\Lambda}_{{\cal R}}}

\begin{document}

\thispagestyle{empty}

\vspace*{28mm}
\begin{center}
{\LARGE\bf The Banks-Zaks expansion \\
\vspace*{4mm}
 in perturbative QCD:  an update}

\vspace{23mm}
{\large P. M. Stevenson}
\vspace{4mm}\\
{\it
T.W. Bonner Laboratory, Department of Physics and Astronomy,\\
Rice University, Houston, TX 77251, USA}

\vspace{27mm}

{\bf Abstract:}

\end{center}

\begin{quote}
The recent QCD calculations of the five-loop $\beta$-function and of 
$R_{e^+e^-}$ to $O(\alpha_s^4)$ provide one more term in the 
Banks-Zaks expansion in $(16 \half - n_{f})$.    
There is no longer any hope that the expansion could extend, even crudely, 
to low $n_f$.  Above $n_f \sim 9$, however, the results appear to be 
reasonably consistent from order to order.

\end{quote}

\newpage

   This Letter is to update earlier work \cite{st,Caveny}, taking into account 
the new results of  Baikov, Chetyrkin, and  K\"uhn for the fifth-order $\beta$ function 
\cite{c4calc} and for $R_{e^+e^-}$ at fourth order \cite{r3calc}.  Contrary to our 
original hopes, the Banks-Zaks (BZ) expansion \cite{bz}-\cite{grun} appears to 
break down around $n_f \sim 9$ or sooner, and does not extrapolate, even crudely, to 
low $n_f$.  

     We write the $\beta$ function in the form:
\BE
\label{eq: beta}
\beta \left( a\right) \equiv \mu \frac{da}{d\mu} = -b a^2 \left( 1 + ca + 
c_{2} a^2 + c_{3} a^3 +c_{4} a^4 +\ldots \right) \,,
\EE
where  $a \equiv  \alpha_{s}/\pi$.  The coefficients, in the $\msbar$ 
scheme, are \cite{bcalc}-\cite{c3calc},\cite{c4calc}:
\begin{eqnarray}
 2b &  =  & 11 - \frac{2}{3} n_f \,,
\nonumber \\
 8 bc & =  & 102 - \frac{38}{3} n_f  \,,
\nonumber \\
 32 b c_2 & = & \frac{2857}{2} - 
\frac{5033}{18} n_f + \frac{325}{54} n_f^2 \,,
\\
 128 b c_3 & = & 
\left( \frac{149753}{6} + 3564 \zeta_3 \right) 
- \left( \frac{1078361}{162} + \frac{6508}{27} \zeta_3 \right) n_f 
\nonumber \\
& \!\!\!\!& \mbox{}
+ \left( \frac{50065}{162} + \frac{6472}{81} \zeta_3 \right) n_f^2 
+ \frac{1093}{729} n_f^3,
\nonumber \\
 512 b c_4  & = & \left( \frac{8157455}{16} + \frac{621885}{2} \zeta_3 - \frac{88209}{2} \zeta_4 - 
  288090 \zeta_5 \right)
  \nonumber \\
& \!\!\!\!& \mbox{}
+ \left( -\frac{336460813}{1944} - \frac{4811164}{81} \zeta_3 + \frac{33935}{6} \zeta_4 + 
\frac{1358995}{27} \zeta_5 \right) n_f 
\nonumber \\ 
& \!\!\!\!& \mbox{}
+  \left( \frac{25960913}{1944} + \frac{698531}{81} \zeta_3 - \frac{10526}{9} \zeta_4 
- \frac{381760}{81} \zeta_5 \right) n_f^2
\nonumber \\
& \!\!\!\!& \mbox{}
+ \left( - \frac{630559}{5832} - \frac{48722}{243} \zeta_3 + \frac{1618}{27} \zeta_4 
+ \frac{460}{9} \zeta_5 \right) n_f^3 + \left( \frac{1205}{2916} - \frac{152}{81} \zeta_3 \right) n_f^4.
\nonumber 
\end{eqnarray}
Here $\zeta_s$ is the Riemann zeta-function and $n_f$ is the number of massless quark flavours.  

    For $n_{f}$ just below $16 \half$, the $\beta$ function has a 
zero at $a^{*} \sim -\frac{1}{c}$, and $a^*$ is asymptotically 
proportional to $(16 \half - n_{f})$.  Its limiting form,
\BE
\label{a0}
a_{0} \equiv \frac{8}{321}\left( 16 \half - n_{f}\right), 
\EE
is the natural expansion parameter for the BZ expansion \cite{st}.  
To proceed, one re-writes all perturbative coefficients, 
eliminating $n_f$ in favour of $a_0$.  The first two $\beta$-function 
coefficients, which are renormalization-scheme (RS) invariant, become:
\BE
b = \frac{107}{8} a_{0},  \quad\quad\quad
c = -\frac{1}{a_{0}} + \frac{19}{4}. 
\EE
Within the class of so-called `regular' schemes \cite{grun,st}, which 
includes $\msbar$, perturbative coefficients have a polynomial dependence 
on $n_f$, and we may write 
\BE
\label{ciexp}
c_{i} = \frac{1}{a_{0}} \left( c_{i,-1} + c_{i,0} a_{0} + c_{i,1} a_{0}^2 
+ \ldots \right).
\EE
The coefficients, in $\msbar$, are collected in the table below.

\vspace*{2mm}

\begin{center}
\begin{tabular}{|rclcr|}
\hline
$c_{1,0}$ &=& $\frac{19}{4}$ &=& 4.75 \\ 
\hline 
$c_{2,-1}$ &$=$& $- \left( \frac{8}{107} \right) 
\left( \frac{37117}{768} \right) $ &=& $-3.61$  \\  
$c_{2,0}$ &=& $\frac{243}{32}$ &=& 7.59 \\ 
$c_{2,1}$ &=& $\left( \frac{107}{8} \right) 
\left(  \frac{325}{192} \right) $ &=& 22.6 \\
\hline
$c_{3,-1}$  &=& $ \left( \frac{8}{107} \right) 
\left( \frac{53981}{1152} + \frac{5335}{32}\zeta_3 \right) $ 
&=& 18.5 \\
$c_{3,0}$ &=& $-\frac{1544327}{13824} - \frac{16171}{288}\zeta_3$ 
&=& $-179$ \\
$c_{3,1}$  &=& $\left( \frac{107}{8} \right) 
\left( \frac{2587}{96} +\frac{809}{144}\zeta_3 \right) $ 
&=& 451 \\
$c_{3,2}$ &=& $ - \left( \frac{107}{8} \right)^2 
\left( \frac{1093}{3456} \right) $ &=& $-56.6$ \\
\hline
$c_{4,-1}$ &=& $\left( \frac{8}{107} \right) 
\left( \frac{1081830511}{663552} + \frac{17251949}{13824} \zeta_3 - 
\frac{191675}{192} \zeta_5 \right) $ &=& $156.7$\\
$c_{4,0}$ &=& $
- \frac{1452057293}{1327104} - \frac{48015}{512} \zeta_4 - 
\frac{4489165}{27648} \zeta_3 + \frac{856625}{2304} \zeta_5  $ &=& $-1005.3$ \\
$c_{4,1}$ &=& $\left( \frac{107}{8} \right) \left(  \frac{33737869}{221184} + \frac{16171}{512} \zeta_4
 - \frac{176837}{2304} \zeta_3 - \frac{88415}{2304} \zeta_5 \right) $  &=& $731.1 $ \\
$c_{4,2}$ &=& $\left( \frac{107}{8} \right)^2 \left(
\frac{471499}{110592} - \frac{809}{256} \zeta_4 + \frac{39409}{2304} \zeta_3 - \frac{345}{128} \zeta_5 
\right) $  &=&  $ 3329.0$ \\
$c_{4,3}$ &=& $ \left( \frac{107}{8} \right)^3 
\left( \frac{1205}{18432} - \frac{19}{64} \zeta_3\right) $ &=& $-697.4 $ \\
\hline
 
\end{tabular}

\vspace*{5mm}
Table 1.  {\it $\beta$-function coefficients in the $\msbar$ scheme.}

\end{center}

\vspace*{3mm}

    The BZ expansion can be applied to any perturbatively calculable
physical quantity of the form:
\BE
\label{eq: R}
{\cal{R}}  = a \left( 1 + r_{1} a + r_{2} a^2 + r_{3} a^3 
+ \ldots\right).  
\EE
For `primary' quantities calculated in a `regular' scheme the coefficients $r_{i}$ are 
polynomials in $n_f$, and hence in $a_{0}$:
\BE 
\label{riexp}
r_i = r_{i,0} + r_{i,1} a_0 + r_{i,2} a_0^2 + \ldots .
\EE
Note that a term $r_{i,j} a_0^p$ or $c_{i,j} a_0^p$ can be assigned a 
degree $i+j-p$, and all terms in any formula must have matching degree.  
[We mention that the same decomposition of coefficients is needed 
in the ``large-$b$'' approximation \cite{largeb,max}, which employs the 
opposite limit ($b \to \infty$), rather than $b = \frac{107}{8} a_0 \to 0$ as here.]  

    The prototypical example is the $e^+e^-$ ratio at a {\it cm} energy $Q$:
\BE
R_{e^{+} e^{-}}(Q) \equiv \frac{\sigma_{\rm tot} \left( e^{+} e^{-} \rightarrow 
{\mbox{\rm hadrons}}\right)}{\sigma \left( e^{+} e^{-} \rightarrow \mu^{+} 
\mu^{-}\right)},
\EE
where, neglecting quark masses, $R_{e^{+} e^{-}} \left( Q \right) = 
3 \Sigma q_{i}^2 \left( 1 + {\cal R}_{e^+e^-}\right)$, and ${\cal R}_{e^+e^-}$ 
has the form of Eq.~(\ref{eq: R}).  [We will drop ``singlet'' terms proportional to 
$(\Sigma q_{i})^2/(3 \Sigma q_{i}^2)$ whose $n_f$ dependence is ambiguous and 
depends on the electric charges assigned to the additional, fictitious quarks.]  
The coefficients, calculated in $\msbar$ with the renormalization scale $\mu$ equal 
to $Q$ \cite{r1calc,r2calc,r3calc}, are collected in the table below.

\vspace*{2mm}

\begin{center}
\begin{tabular}{|rclcr|}

\hline
$r_{1,0}$ &=& $\frac{1}{12}$ &=& $0.0833$ \\
$r_{1,1}$ &=& $\left( \frac{107}{8} \right) 
\left( \frac{11}{4} - 2 \zeta_3 \right) $ &=& $4.63$ \\  
\hline
$r_{2,0}$ &=& $-\frac{12521}{288} + 13 \zeta_3$ &=& $-27.85$ \\ 
$r_{2,1}$ &=& $ \left( \frac{107}{8} \right) 
\left( \frac{401}{24} - \frac{53}{3}\zeta_3 + \frac{25}{3}\zeta_5 \right) $ 
&=& 55.0 \\
$r_{2,2}$ &=& $ \left( \frac{107}{8} \right)^2 
\left( \frac{151}{18} - \frac{19}{3}\zeta_3 - \frac{1}{2} \zeta_2 \right) $ 
&=& $-8.34$ \\
\hline
$r_{3,0}$ &=& $ - \frac{3963761}{20736} + \frac{677833}{3456} \zeta_3 - 
\frac{275}{24} \zeta_5  $ &=&  $ 32.73 $ \\
$r_{3,1}$ &=& $ \left( \frac{107}{8} \right)  \left( -\frac{38969}{128} + \frac{535}{32} \zeta_2
+ \frac{6907}{96} \zeta_3 + \frac{165}{2} \zeta_3^2 + \frac{9595}{144} \zeta_5 - \frac{665}{24} \zeta_7 \right)  $ 
&=&  $ -402.6 $ \\
$r_{3,2}$ &=& $ \left( \frac{107}{8} \right)^2  \left( \frac{236089}{1728} - \frac{97}{16} \zeta_2 
- \frac{13859}{96} \zeta_3 + \frac{15}{2} \zeta_3^2 + \frac{445}{12} \zeta_5 \right)   $  &=&  $ 430.9 $  \\
$r_{3,3}$ &=& $ \left( \frac{107}{8} \right)^3  \left( \frac{6131}{216} - \frac{33}{8} \zeta_2 
- \frac{203}{12} \zeta_3 + 3 \zeta_2 \zeta_3 - \frac{15}{2} \zeta_5   \right)    
$  &=&  $ -1390.0 $  \\
\hline

\end{tabular}

\vspace*{5mm}
Table 2.  {\it Coefficients in ${\cal R}_{e^+e^-}$ in the $\msbar$($\mu=Q$) scheme.}

\end{center}

\vspace*{3mm}

     The fixed-point condition $\beta(a^{*}) = 0$ always has a solution 
as a power series in $a_0$:
\BE
\label{astar}
a^{*} = a_{0} \left( 1 + v_1 a_{0} + v_2 a_{0}^{2} + 
v_3 a_{0}^{3} + \ldots \right).
\EE
A straightforward calculation yields:
\begin{eqnarray}
\label{eq: AA}
v_1 &=& c_{1,0} +c_{2,-1},\nonumber \\
v_2 &=& (c_{1,0}+2c_{2,-1})(c_{1,0} +c_{2,-1})+c_{2,0}+c_{3,-1}, \\  
v_3 &=&
c_{1,0}^3 + 6 c_{1,0}^2 c_{2,-1} + 
c_{1,0}(3 c_{2,0} + 4 c_{3,-1} + 10 c_{2,-1}^2)
\nonumber \\
& & {\mbox{}} + c_{2,-1}(4 c_{2,0} + 5 c_{3,-1}) + 5 c_{2,-1}^3 + c_{2,1} +
c_{3,0} + c_{4,-1} \,. \nonumber
\end{eqnarray}
Numerically, $v_1 =1.1366$, $v_2 =23.27$, $v_3 =18.10 $, 
in the $\msbar$ scheme.  Since $a^*$ is RS dependent, the good or bad convergence 
of this series need not concern us.

     A physical quantity ${\cal R}$ also has an infrared limit, ${\cal R}^*$, given by a 
power series in $a_{0}$.  Substituting $a=a^*$ from Eq.~(\ref{astar}) into the 
perturbative expansion of ${\cal R}$ and re-expanding in powers of $a_{0}$ yields
\BE
\label{eq: RR}
{\cal R}^{*} = a_{0} \left( 1 + w_1 a_{0} + w_2 a_{0}^{2} + 
w_3 a_{0}^{3} + \ldots  \right),
\EE
where
\begin{eqnarray}
w_1 &=& v_1 +r_{1,0},\nonumber \\
w_2 &=& v_2 + 2r_{1,0} v_1 + r_{2,0} + r_{1,1}, \\
w_3 &=& v_3 + (2 v_2 + v_1^2) r_{1,0} + v_1 (2r_{1,1} + 3 r_{2,0})
+ r_{2,1}+ r_{3,0}.\nonumber 
\end{eqnarray}
These coefficients are RS independent.  For the $e^+e^-$ case they are
\begin{eqnarray}
w_1 &=&  \frac{4177}{2^5 (107)}, \nonumber \\
w_2 &=&  \frac{31250575}{2^9 3 (107)^2} - \frac{275}{2(107)} \zeta_3, \\
w_3 &=&  \frac{2177185161509}{2^{15} 3^2 (107)^3} - 
\frac{4232749}{2^6 (107)^2} \zeta_3 
+ \frac{65125}{2^3 3 (107)} \zeta_5.  \nonumber
\end{eqnarray}
Numerically we find
\BE 
\label{ree}
{\cal R}_{e^+e^-}^* = a_0 \left( 1 + 1.22 a_0 + 0.23 a_0^2 + 25.38 a_0^3 + \ldots \right).
\EE
While the first three terms raise hopes for a well-behaved series, those hopes are dashed by the 
last term.  See Fig.~1.


\begin{figure}[!hbt]

\centering
\includegraphics[width=0.65 \textwidth]{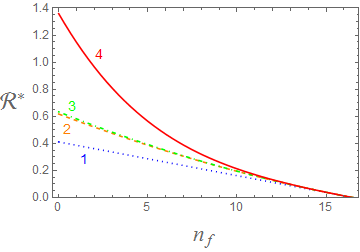}
\vspace*{-2mm}
\begin{quote}

{\setlength{\baselineskip}{0.8\baselineskip} 
Fig.~1. \,  ${\cal R}^*$ as a function of $n_f$ in the BZ expansion.  The curves for $1^{st}$ to 
$4^{th}$ order are shown dotted, dashed, dot-dashed, and solid.
 \par}
 
\end{quote}
\end{figure}

    A formulation of the BZ expansion for quantities at a general $Q$ 
was derived in Ref.~\cite{st}.  First, we write the integrated 
$\beta$-function equation in the form 
\BE
\label{eq: int} 
b \ln \left( \frac{\mu}{\tilde{\Lambda}} \right) = 
\frac{1}{a} + c \ln (\mid\!c\!\mid\! a) + 
\int_0^a dx \left( \frac{b}{\beta(x)} + \frac{1}{x^2} - \frac{c}{x} \right) .
\EE
This form, more convenient for $c$ negative, is completely equivalent to our 
previous definition of the $\tilde{\Lambda}$ parameter \cite{OPT,Caveny}.
We use a tilde to distinguish $\tilde{\Lambda}$ from the conventional definition 
of the $\Lambda$ parameter \cite{bbdm}.  The relation is 
$\ln(\Lambda/\tilde{\Lambda}) = (c/b) \ln(2 \! \mid \!\! c \! \mid \!\! /b)$.  
The two definitions are not dissimilar for small $n_f$, but they become infinitely 
different as $n_f \to 16 \half$.  In the BZ-expansion context the use of 
$\tilde{\Lambda}$ is much more convenient.

    As explained in Ref. \cite{st}, it is convenient to put the $\beta$ 
function into the form 
\BE
\label{betaf}
\frac{b}{\beta\left(x\right)} = - \frac{1}{x^2} + \frac{c}{x} - 
\frac{b}{\gamma^* \left(a^{*} - x\right)} +  H\left(x\right).
\EE
where $\gamma^{*}$ is 
the slope of the $\beta$ function at the fixed point:
\BE
\label{gamstar}
\gamma^{*} \equiv \left. \frac{d \beta\left(x\right)}{dx}\right|_{x=a^{*}} 
= -b a^{*} 
\left( 1 + 2 ca^{*} + 3 c_{2} {a^{*}}^{2}+ 4 c_{3} {a^{*}}^{3} +\ldots \right).
\EE
As discussed below, $\gamma^*$ can be obtained as a series in $a_0$.  
The remainder function $H(x)$ can be expanded as a power series, 
$H_0 + H_1 x + \ldots$, whose coefficients are of order $a_0$.  

    One now inserts (\ref{betaf}) into (\ref{eq: int}) and performs the 
integration.  One can then eliminate $a$ and $a^{*}$ in favour of 
${\cal R}$ and ${\cal R}^{*}$.  In fact, since the result must be RS 
invariant, one can --- without loss of generality --- short-cut this 
step by utilizing the ``effective-charge'' RS \cite{Grunberg} in which 
$a \equiv {\cal R}$.  In $n^{th}$ order of the BZ expansion this leads to 
the formula 
\cite{st}:
\BE 
\label{eq: YY}
\boldsymbol{\rho}_{1}(Q) = \frac{1}{\cal R} + 
c\ln\left(\left|c\right| {\cal R} \right) + 
\frac{b}{\gamma^{*(n)}} \ln\left(1- \frac{{\cal R}}{{\cal R}^{*(n)}} \right) 
+ \sum_{i=0}^{n-4} \frac{H_i^{\rm \scriptscriptstyle{(EC)}} {\cal R}^{i+1}}{i+1}.
\EE
On the left-hand side, $\boldsymbol{\rho}_1(Q)$ is the 
RS invariant 
\BE
\label{eq: X}
\boldsymbol{\rho}_{1}(Q) \equiv b\ln\left( \frac{\mu}{\tilde{\Lambda}}\right) - r_{1} 
\equiv b \ln\left( \frac{Q}{\lamR} \right),
\EE
where $\lamR$ is a characteristic scale specific to 
the particular physical quantity ${\cal R}$.  It is related to the 
$\tilde{\Lambda}$ parameter of some reference scheme (eg.~$\msbar$) by 
an exactly calculable factor $\exp(r_1({\scriptstyle{\mu=Q}})/b)$ involving the 
$r_1$ coefficient in that scheme, evaluated at $\mu=Q$.  On the right-hand side 
the terms involving the $H_i^{\rm \scriptscriptstyle{(EC)}}$ coefficients of the 
effective-charge scheme are only relevant in fourth order and beyond.  
Thus, for the first three orders the equation takes the same form, just with 
the parameters $\gamma^*$ and ${\cal R}^*$ approximated to the 
appropriate order.  At $4^{th}$ order there is an extra term 
$H_0^{\rm \scriptscriptstyle{(EC)}} {\cal R}$, with 
$H_0^{\rm \scriptscriptstyle{(EC)}}=H_{0,1}^{\rm \scriptscriptstyle{(EC)}} a_0+O(a_0^2)$, 
where  
\begin{eqnarray}
H_{0,1}^{\rm \scriptscriptstyle{(EC)}} & = &  \rho_{4,-1}+2 \rho_{2,-1}\rho_{3,-1} +\rho_{2,-1}^3  ,
 \nonumber \\
   & = &  c_{4,-1} + c_{2,-1} (2 c_{3,-1} + r_{1,0}^2 - r_{2,0}) 
   + c_{2,-1}^3  \nonumber \\
  &  & {\mbox{}}  - c_{2,-1}^2 r_{1,0} - c_{3,-1} r_{1,0} - r_{1,0}^3 
  + 2 r_{1,0} r_{2,0} - r_{3,0},  \\
   & & \nonumber \\
  & = &  \frac{243227350299721}{2^{15} 3^4 (107)^3} - 
  \frac{5729638277}{2^7 3^3 (107)^2} \zeta_3 - 
  \frac{81125}{2^2 3 (107)} \zeta_5  
  \quad \simeq -164.8  \nonumber
\end{eqnarray}
(in the first line, the $\rho_{i,j}$ are the $\beta$-function coefficients of the EC scheme).

The BZ expansion 
for $\gamma^*$ is obtained straightforwardly by substituting the 
expansion of $a^*$ (Eqs. (\ref{astar}) and (\ref{eq: AA})) into 
(\ref{gamstar}).  This gives:
\BE
\gamma^{*} =b a_{0}\left( 1 + g_1 a_{0} + 
g_2 {a_{0}}^{2} + g_3 {a_{0}}^{3} + \ldots \right) ,
\EE
with
\begin{eqnarray}
g_1 &=&  c_{1,0},\nonumber 
\\
g_2 &=& {c_{1,0}}^{2} - {c_{2,-1}}^{2} -c_{3,-1} 
, \\
g_3 &=&{c_{1,0}}^{3} - {4c_{2,-1}}^3 
- 5 {c_{1,0}}{c_{2,-1}}^{2} - 4 {c_{1,0}} c_{3,-1}
\nonumber \\
& & \mbox{} - 2 c_{2,-1} c_{2,0} - 6 c_{2,-1} c_{3,-1} - c_{3,0} - 2c_{4,-1}.
\nonumber
\end{eqnarray}
It is noteworthy that certain terms of degree $n$ are absent in $g_n$:  
$g_1$ does not contain $c_{2,-1}$; $g_2$ does not contain $c_{2,0}$ or 
$c_{2,-1} c_{1,0}$; and $g_3$ does not contain $c_{2,1}$ or $c_{2,0} c_{1,0}$ 
or $c_{2,-1} {c_{1,0}}^2$.  

The values of these invariants are 
\begin{eqnarray}
g_1 &=&  \frac{19}{4} , \nonumber 
\\
g_2 &=&  \frac{633325687}{2^{10} 3^2 (107)^2}
- \frac{5335}{2^2 (107)} \zeta_3 
, \\
g_3 &=&   -\frac{66670528901419}{2^{13} 3^4 (107)^3}
 - \frac{1920043907}{2^6 3^3 (107)^2} \zeta_3 
+ \frac{191675}{2^2 3 (107)} \zeta_5 .
\nonumber
\end{eqnarray}
Numerically the 
$\gamma^*$ series is:
\BE
\label{gamser}
\gamma^{*} = b a_{0}\left( 1 + 4.75 a_{0} - 8.98 {a_{0}}^{2} 
- 43.89 {a_{0}}^3 + \ldots \right) .
\EE
The results, at different orders, are shown in Fig.~2.


\begin{figure}[!hbt]

\centering
\includegraphics[width=0.65 \textwidth]{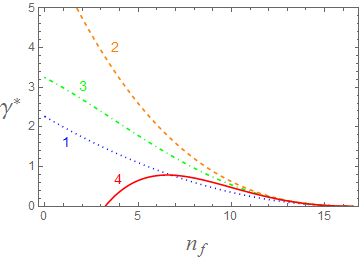}
\vspace*{-2mm}
\begin{quote}

{\setlength{\baselineskip}{0.8\baselineskip} 
Fig.~2. \,  $\gamma^*$ as a function of $n_f$ in the BZ expansion.  The curves 
for $1^{st}$ to $4^{th}$ order are shown dotted, dashed, dot-dashed, and solid.
 \par}
 
\end{quote}
\end{figure}

      Note that $\gamma^*$ is the `critical exponent' in the relation 
${\cal R}^* - {\cal R} \propto Q^{\gamma^{*}}$ that describes how ${\cal R}$ 
approaches ${\cal R}^{*}$ as $Q \rightarrow 0$.  ($\gamma^*$ is the infrared limit of 
an RS-invariant `effective exponent' 
$\gamma(Q) \equiv  1+Q \, \frac{d^2{\cal R}}{dQ^2} \Big/  \frac{d{\cal R}}{dQ}
=   \frac{d \beta}{da} + 
\beta(a) \frac{d^2 {\cal R}}{d a^2} \Big/ \frac{d{\cal R}}{d a} $ 
\cite{effexp}.) 
As pointed out by Grunberg \cite{grun}, the $g_n$ coefficients are RS invariants 
and are universal, in the sense that they are not specific to a
particular physical quantity ${\cal R}$.  

    Numerically inverting Eq. (\ref{eq: YY}) provides ${\cal R}$ as a 
function of $Q$.  In the BZ region, $n_f \gtrsim 9$, the resulting ${\cal R}(Q)$ has 
the general form sketched in Fig.~3.  
At large $Q$ the result naturally agrees with ordinary perturbation theory to the 
corresponding order.  For $Q \sim \lamR $ there is a large ``sloping plateau'' region, 
and at ultra-low energies there is a ``spike'' reaching up to ${\cal R}^{*(n)}$.

\begin{figure}[!hbt]

\centering
\includegraphics[width=0.50 \textwidth]{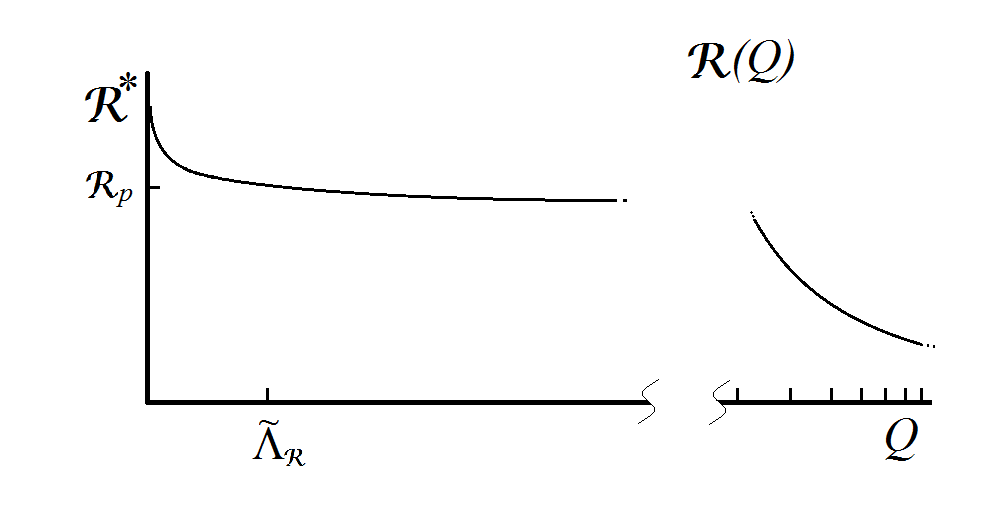}
\vspace*{-2mm}
\begin{quote}

{\setlength{\baselineskip}{0.8\baselineskip} 
Fig.~3.  Typical appearance of ${\cal R}$ as a function of $Q$ in the BZ region 
($n_f \gtrsim 9$) showing the ``spike'' at very low energies, the ``sloping plateau'' region, 
and the slow approach to asymptotic freedom at very high energies (this region is shown 
on a log scale).  The plateau value ${\cal R}_p$ is generally about $0.8$ times 
${\cal R}^*$ but depends on $n_f$ and the BZ-expansion order.
 \par}
 
\end{quote}
\end{figure}

    We conclude by showing, in Fig.~4, a comparison of the $4^{th}$ order BZ expansion with 
the ${\cal R}_{e^+e^-}^*$ results of Ref.~\cite{unfixed} in optimized perturbation theory 
(OPT) \cite{OPT} and in the EC scheme \cite{Grunberg} to order $\alpha_s^4$.  Contrary to 
the conjecture of Refs.~\cite{st,Caveny}, it now appears that the ``freezing'' behaviour of 
${\cal R}_{e^+e^-}^*$ at low $n_f$ \cite{CKL,lowen} is not an extrapolation from the BZ 
region, but a distinct phenomenon.\footnote{
At low $n_f$ it appears that different physical quantities may have rather different infrared 
behaviours \cite{GardiK}, unlike the BZ region where there is a high degree of universality.
}
At low $n_f$ one finds that $\gamma^*$ is around $2$ or $3$, so that ${\cal R}$ ``freezes,'' 
becoming nearly constant in the infrared region, while it falls rapidly around $Q \simeq \lamR$.   
In the BZ region, by contrast, $\gamma^*$ is small ($\lesssim 1$), resulting in the infrared 
``spike'' of Fig.~3 and the sloping plateau around $Q \simeq \lamR$.

    The OPT and EC results in Fig.~4 agree remarkably well at both low and high $n_f$.  
In the intermediate region $7 \lesssim n_f  \lesssim 13$ they actually differ only at the very 
lowest energies, because OPT indicates a much more dramatic ``spike'' in the infrared, of 
very uncertain size --- it could well be even bigger than predicted.  This is because the 
infrared limit in OPT here does not result from a fixed point but from an ``unfixed point'' 
and a ``pinch mechanism'' that leads to 
$({\cal R}^\star -{\cal R})  \propto 1/\!\mid\!\ln Q \!\mid^2$, corresponding 
to $\gamma^*=0$.  For details, see Ref.~\cite{unfixed}.


\begin{figure}[tb]

\centering
\includegraphics[width=0.85 \textwidth]{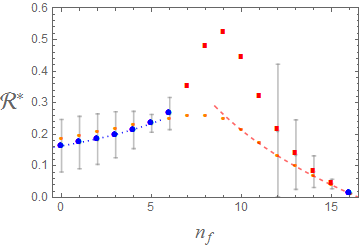}
\vspace*{-2mm}
\begin{quote}

{\setlength{\baselineskip}{0.8\baselineskip} 
Fig.~4. \,   ${\cal R}^*$ as a function of $n_f$ in the $4^{th}$ order 
BZ expansion (dashed curve) compared with OPT (large points) and  
EC (small points) results from Ref.~\cite{unfixed}.  The OPT points 
are shown as blue circles when they arise from a fixed point and as red squares 
when they arise from an ``unfixed point.''  Error bars indicate the estimated 
uncertainty in the OPT results.  (They are not shown for $n_f=7, \ldots, 11$, 
where they would extend well beyond the bounds of the plot.)  
The dotted blue curve represents ${\cal R}^* = 0.9/b$, a purely speculative 
guess at the large-$b$ form.
 \par}
 
\end{quote}
\end{figure}

\vspace*{7mm}

 {\bf Acknowledgment}:  This note is partly based on unpublished work done 
 in collaboration with Scott Caveny.


\newpage


\begin{thebibliography}{99}

\bibitem{st} 
  P.~M. Stevenson, Phys. Lett. B {\bf 331}, 187 (1994).
  
\bibitem{Caveny} 
  S. A. Caveny and P. M. Stevenson, hep-ph/9705319 (unpublished).
  
\bibitem{c4calc}
  P. A. Baikov, K. G. Chetyrkin, and  J. H. K\"uhn, arXiv 1606.08659 [hep-ph].  
  
\bibitem{r3calc}
  P. A. Baikov, K. G. Chetyrkin, J. H. K\"uhn, and J. Rittinger, 
  arXiv:1206:1288 (2012);  P. A. Baikov, K. G. Chetyrkin, and J. H. K\"uhn, 
  Phys. Rev. Lett. {\bf 101}, 012002 (2008).
  
\bibitem{bz}
  T. Banks and A. Zaks, Nucl. Phys. B {\bf 196}, 189 (1982).
  
\bibitem{white}
  A.~R. White, Phys. Rev. D {\bf 29}, 1435 (1984); in {\it Hadronic Matter in
  Collision}, edited by J. Rafelski (World Scientific, 1989); Int. J. Mod. 
  Phys. A {\bf 8}, 4755 (1993).
  
\bibitem{grun}
  G. Grunberg, Phys. Rev. D {\bf 46}, 2228 (1992).

\bibitem{bcalc}
  H. D. Politzer, Phys. Rev. Lett. {\bf 30}, 1346 (1973);  
  D. J. Gross and F. Wilczek, {\it ibid.} {\bf 30}, 1343 (1973); 
  G. 't Hooft, report at the Marseille Conference Yang-Mills Fields, 1972.

\bibitem{ccalc}
  D. R. T. Jones, Nucl. Phys. B {\bf 75}, 531 (1974); 
  W. Caswell, Phys. Rev. Lett. {\bf 33}, 244 (1974);
  E. S. Egorian and O. V. Tarasov, Theor. Mat. Fiz. {\bf 41}, 26 (1979).

\bibitem{c2calc}
   O. V. Tarasov, A. A. Vladimirov, and A. Yu. Zharkov, Phys. Lett. B 
  {\bf 93}, 429 (1980);
  S. A. Larin and J. A. M. Vermaseren, Phys. Lett. B {\bf 303}, 334 (1993).

\bibitem{c3calc}
  T. van Ritbergen, J. A. M. Vermaseren, and S. A. Larin, Phys. Lett. B 
  {\bf 400}, 379 (1997).
  
\bibitem{largeb}
  M. Beneke, Nucl. Phys. B {\bf 405}, 424 (1993);
  D.~J. Broadhurst, Z. Phys. C {\bf 58}, 339 (1993);
  K. Van Acoleyen and H. Verschelde, Phys. Rev. D {\bf 69} 125006 (2004). 

\bibitem{max}
  C.~N. Lovett-Turner and C.~J. Maxwell, Nucl. Phys. B {\bf 432}, 147 (1994);
{\it ibid} {\bf 452}, 188 (1995); C.~J. Maxwell and D.~G. Tonge, {\it ibid} 
{\bf 481}, 681 (1996); P. M. Brooks and C. J. Maxwell, Phys. Rev. D {\bf 74}, 065012 (2006).

 \bibitem{r1calc}
  K. G. Chetyrkin, A. L. Kataev, and F. V. Tkachov, Phys. Lett. B {\bf 85},
  277 (1979);  M. Dine and J. Sapirstein, Phys. Rev. Lett. {\bf 43}, 668 
  (1979); W. Celmaster and R. J. Gonsalves, Phys. Rev. D {\bf 21}, 3112 (1980).
  
\bibitem{r2calc} 
  L. R. Surguladze and M. A. Samuel, Phys. Rev. Lett. {\bf 66},
  560 (1991);  {\it ibid} 2416 (E); 
  S. G. Gorishny, A. L. Kataev, and S. A. Larin, Phys. Lett. B {\bf 259}, 144 (1991).  

\bibitem{OPT}
  P. M. Stevenson, Phys. Rev. {\bf D23}, 2916 (1981). 

\bibitem{bbdm}
  W.~A. Bardeen, A.~J. Buras, D.~W. Duke, and T. Muta, Phys. Rev. D 
{\bf 18}, 3998 (1978).

\bibitem{Grunberg}
  G. Grunberg, Phys. Rev. D {\bf 29}, 2315 (1984); A. Dhar and V. Gupta, 
  Phys. Rev. D {\bf 29}, 2822 (1984).

\bibitem{effexp}
  P. M. Stevenson, arXiv:~1606.06951 [hep-ph].

\bibitem{unfixed}
  P. M. Stevenson, Nucl. Phys. B {\bf 875}, 63 (2013).  

\bibitem{CKL}
  J. Ch\'yla, A. Kataev, and S. A. Larin, Phys. Lett. B {\bf 267}, 269 (1991).

\bibitem{lowen}  
  A. C. Mattingly and P. M. Stevenson, Phys. Rev. Lett. {\bf 69}, 1320 (1992); 
  Phys. Rev. D {\bf 49}, 437 (1994); 
  P. M. Stevenson, Nucl. Phys. B {\bf 868}, 38 (2013).

\bibitem{GardiK}
   E. Gardi and M. Karliner, Nucl. Phys. B {\bf 529}, 383 (1998).

\end{thebibliography}
\end{document}